\documentclass[12pt]{iopart}

\usepackage{iopams}
\usepackage[amssymb]{SIunits}
\usepackage{graphicx}
\begin{document}

\title[Transport in a three-terminal graphene quantum dot in the multi-level regime]{Transport in a three-terminal graphene quantum dot in the multi-level regime}

\author{A. Jacobsen, P. Simonet, K. Ensslin and T. Ihn}
\address{Solid State Physics
Laboratory, ETH Zurich, 8093 Zurich, Switzerland}
\ead{arnhildj@phys.ethz.ch}


\begin{abstract}
We investigate transport in a  three-terminal graphene quantum dot. All nine elements of the conductance matrix have been independently measured. In the Coulomb blockade regime accurate measurements of individual conductance resonances reveal slightly different resonance energies depending on which pair of leads is used for probing. Rapid changes in the tunneling coupling between the leads and the dot due to localized states in the constrictions has been excluded by tuning the difference in resonance energies using in-plane gates which couple preferentially to individual constrictions. The interpretation of the different resonance energies is then based on the presence of a number of levels in the dot with an energy spacing of the order of the measurement temperature. In this multi-level transport regime the three-terminal device offers the opportunity to sense if the individual levels couple with different strengths to the different leads. This in turn gives qualitative insight into the spatial profile of the corresponding quantum dot wave functions.
\end{abstract}


\maketitle

\tableofcontents


\section{Introduction}

Graphene nanostructures are believed to have potential applications in both conventional
electronics and solid–state quantum information processing. In particular, graphene quantum dots might be promising for spin qubits due to their predicted long spin life times. \cite{trauzettel2007}

As a consequence of the gapless band structure charge carriers cannot be electrostatically confined in graphene. However, by cutting graphene into narrow ribbons a so-called transport gap is opened where the current is suppressed around the charge neutrality point. \cite{han2007,chen2007,stampfer2009,molitor2009,han2010,droscher2011} By using short and narrow constrictions as tunnel barriers more complicated nanodevices like quantum dots have been successfully created. This has led to a number of experiments where, for example, excited states have been observed in single \cite{schnez2009} and double quantum dots \cite{molitor2010,liu2010,volk2011}, spin states have been investigated \cite{guttinger2010} and the electron-hole crossover has been studied \cite{guttinger2009}. In addition, the modulation of transport through graphene quantum dots due to localized states in the constrictions have been investigated in several studies. \cite{stampfer2008,schnez2010,guttinger2011} Still, there are open questions concerning the detailed influence of constriction localizations on transport on small energy scales.

The current through a two-terminal quantum dot does not give access to the
individual coupling strengths between the dot and each lead. However, if a dot in the single-level tunneling regime of the Coulomb blockade is connected to three or more leads the individual tunnel coupling constants between each lead and the quantum dot can be determined from measurements of the conductance matrix of the system. \cite{leturcq2004}

Following the approach of Ref.\,\cite{leturcq2004}, we here investigate transport in a three-terminal graphene quantum dot in the multi-level regime. The three terminals offer the possibility for fast and convenient probing of the conductances of each lead, thereby providing further insight into how localized states in the constrictions affect transport through the dot. In addition, being in the multi-level regime gives us the unique chance to experimentally observe how different leads couple with different strengths to different dot states.


\section{Sample and experimental methods}
\label{exp method}

\begin{figure}
  \begin{center}
    \includegraphics[width=0.6\textwidth]{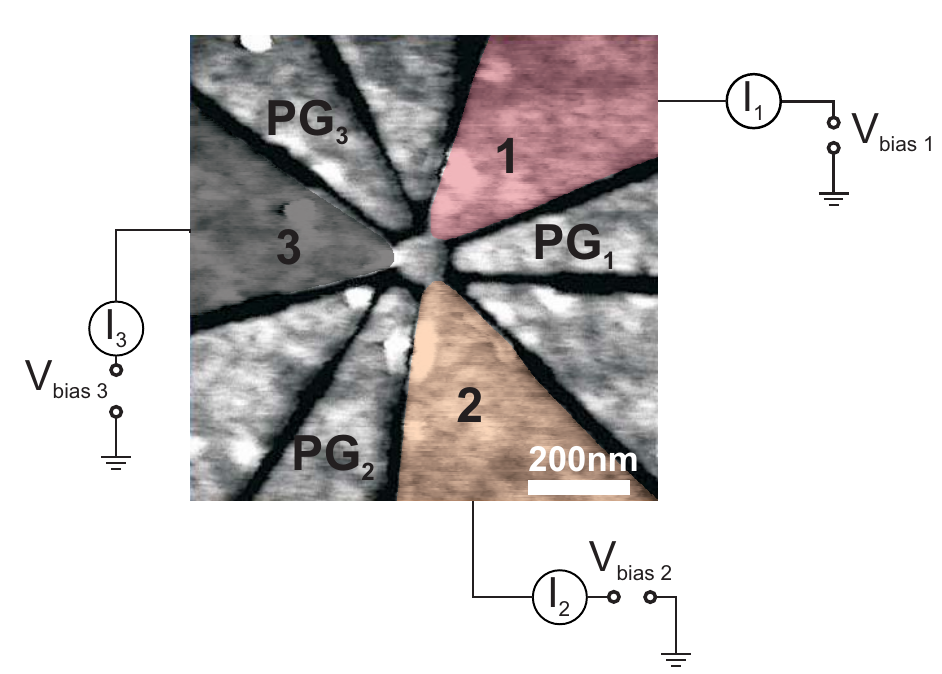}
    \caption{(a) Scanning force micrograph image of the measured
    quantum dot with a sketch of the measurement setup. Leads 1, 2 and 3 are highlighted in red, orange and
    black, respectively. Three plunger gates are used to tune the dot in addition to the global back gate. To each lead a bias voltage can be applied and the current flowing can be measured.
    }
    \label{fig1}
  \end{center}
\end{figure}

Single layer graphene flakes were exfoliated from natural graphite,
deposited onto a highly doped Silicon substrate covered by 285\,nm
thermal silicon dioxide, and identified using Raman spectroscopy \cite{ferrari2006,graf2007} and
light microscopy. In a first electron beam lithography (EBL) step, followed by metal deposition of 5\,nm titanium, 45\,nm gold and lift-off, the Ohmic contacts were added to the flake. The structure is then patterned by a second EBL step followed by reactive ion etching with argon and oxygen (for a detailed description of similar fabrication see Ref.\,\cite{guttinger2009b}).

A scanning force micrograph (SFM) image of the measured quantum dot
is depicted in Fig.\,\ref{fig1}. The quantum dot is connected to
three leads, labeled 1, 2 and 3, through narrow constrictions.
From the SFM image the diameter of the dot is determined to be 110\,nm and the width of the constrictions is found to be 40\,nm. In
addition to the global silicon back gate (BG) three in-plane plunger
gates, PG$_\mathrm{1}$, PG$_\mathrm{2}$ and PG$_\mathrm{3}$, are
used to tune the dot and the constrictions. The
remaining three in-plane gates influence the transport
through the dot  only weakly and are therefore not used.

In Fig.\,\ref{fig1} we additionally sketch the measurement setup. In all
measurements presented in this study a DC bias voltage is applied to
one of the three leads while the other two leads are grounded. The
currents through the three leads are measured simultaneously using
current-voltage converters. All measurements are performed at 1.7\,K unless stated
otherwise.



\section{Results and discussion}


\subsection{Device characterization}

\begin{figure}
  \begin{center}
    \includegraphics[width=1\textwidth]{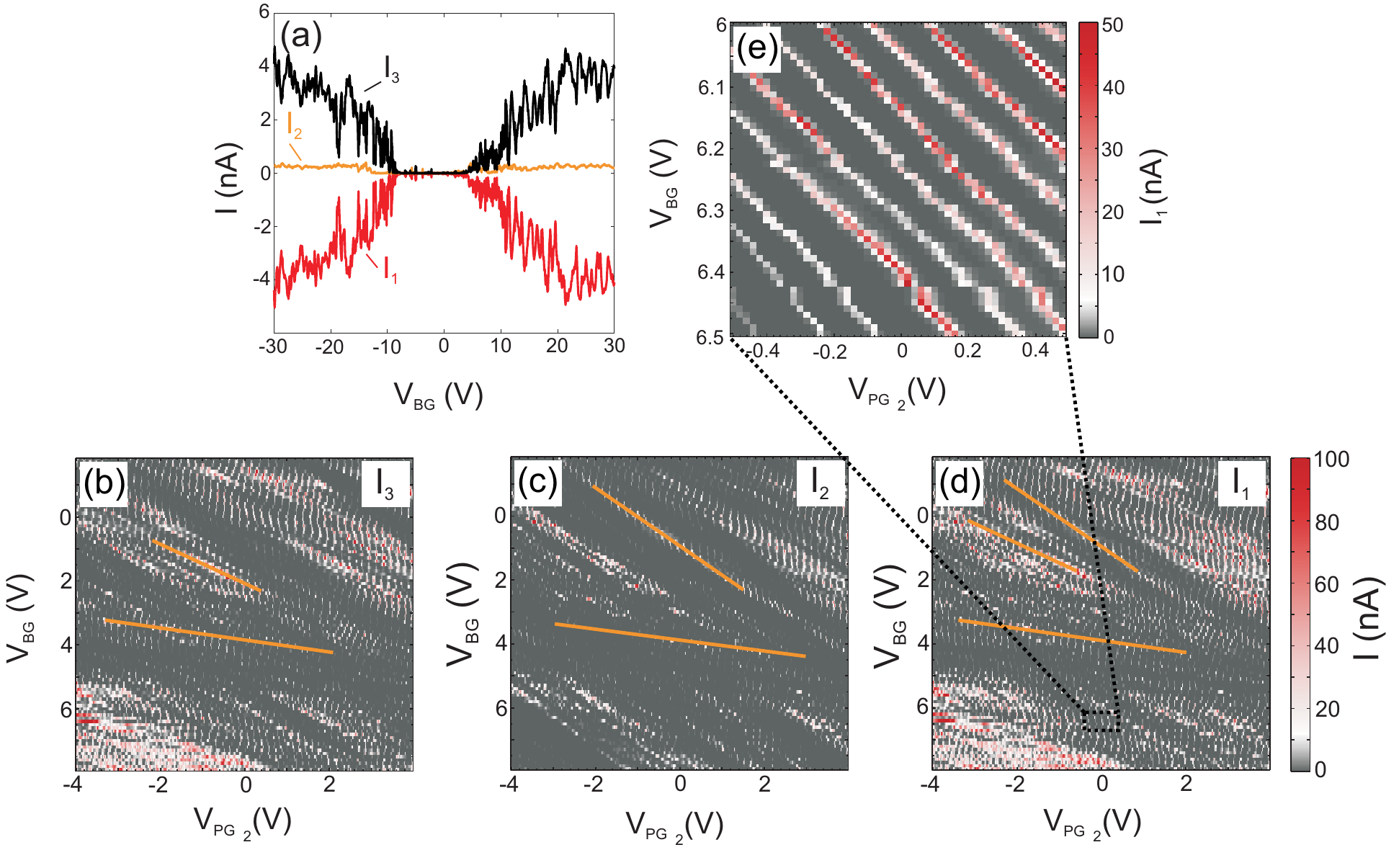}
    \caption{(a) Current through the three leads as a function of back gate
    voltage for a large backgate voltage range. A 1\,mV bias voltage is applied
    to lead 1, while leads 2 and 3 were grounded. A transport gap of $\approx 12$\,V in back gate can be seen. (b)-(d) I$_3$, I$_2$ and I$_1$
    as a function of V$_\mathrm{BG}$ and V$_\mathrm{PG_1}$ for large gate voltage
    ranges. The broad diagonal lines are due to resonances in the constrictions. (e) A zoom of (d) where the narrow diagonal lines corresponds to  Coulomb peaks.}
    \label{fig2}
  \end{center}
\end{figure}

Fig.\,\ref{fig2}(a) shows the currents through the dot for a large
range in V$_\mathrm{BG}$. Curves labeled $I_1$, $I_2$ and $I_3$ correspond to the current measured in lead 1, 2 and 3 respectively. For this measurement 1\,mV bias voltage is applied to lead 1, while leads 2 and 3 are grounded. We use the convention that negative currents flow from the leads into the quantum dot, while positive currents flow from the quantum dot into the leads.

Around the charge neutrality point a region of $\approx12$\,V of
suppressed current, corresponding to the transport gap \cite{stampfer2009}, can be seen. Within this region Coulomb
blockade is observed. From Coulomb diamond measurements we determine the charging energy of the quantum dot to be
8-15\,meV (not shown).

It can also be seen that constriction 2 is generally more closed than 1 and 3. For high charge carrier densities $I_2$ is less than 10$\%$ of $I_1$ (the total current). This asymmetry is also present in the regime of Coulomb blockade where the current flowing through constriction 2 is often too small to be measured and the quantum dot is effectively a two-terminal dot. Still, it is possible
to find regimes where the current contributions from the three leads
are comparable and in the following sections we will focus on one of these regimes.

\begin{table}
\caption{\label{leverarms}Relative lever arms $\alpha_\mathrm{PG}/\alpha_\mathrm{BG}$ for the in-plane gates
with respect to the dot and the three constrictions.}
    \begin{tabular*}{\textwidth}{@{}l*{15}{@{\extracolsep{0pt plus
    12pt}}l}} \br
        & $\alpha_\mathrm{PG}^\mathrm{QD}/\alpha_\mathrm{BG}^\mathrm{QD}$ & $\alpha_\mathrm{PG}^\mathrm{Constr.1}/\alpha_\mathrm{BG}^\mathrm{Constr. 1}$ & $\alpha_\mathrm{PG}^\mathrm{Constr. 2}/\alpha_\mathrm{BG}^\mathrm{Constr. 2}$ & $\alpha_\mathrm{PG}^\mathrm{Constr. 3}/\alpha_\mathrm{BG}^\mathrm{Constr. 3}$\\
        \mr
        PG$_1$ & 0.59 & 1.15 & 0.68 & 0.25\\
        PG$_2$ & 0.50 & 0.13 & 0.88 & 0.65\\
        PG$_3$ & 0.58 & 0.65 & 0.13 & 1.15\\
        \br
    \end{tabular*}
\end{table}

In order to characterize the device further we measure the current
through the three leads as a function of $V_\mathrm{BG}$ and each of the three
plunger gates on both a small and a large voltage scale, and determine the plunger gate lever arms relative to the back gate lever arms $\alpha_\mathrm{PG}/\alpha_\mathrm{BG}$ with respect to the dot and each of the three constrictions. As an example Fig.\,\ref{fig2}(b)-(d) shows $I_1$, $I_2$
and $I_3$ respectively, as a function of $V_\mathrm{BG}$ and $V_{\mathrm{PG}_2}$. A voltage of 1\,mV is applied to lead 1 and $I_1$, $I_2$ and $I_3$ are measured simultaneously. The gates are swept over a large voltage range and the broad diagonal lines that are visible are attributed to resonances in the constrictions. \cite{stampfer2008} In Fig.\,\ref{fig2}(d) three different
slopes (marked by orange lines) can be identified, while only two different slopes can be
found in Fig.\,\ref{fig2}(b) and (c). Since lead 1 is biased all charge carriers flowing through the dot have to tunnel through constriction 1. As a result, resonances originating from localized states in constriction 1 are seen in all three currents. On the other hand, a resonant state in lead 2 or 3 will enhance the current only in this specific lead, with the consequence that only the current through this specific lead (and the biased lead) is enhanced. From this measurement we can therefore assign one slope to states in each constriction and subsequently determine $\alpha_{\mathrm{PG}_2}/\alpha_\mathrm{BG}$ for all three constrictions. Complementary measurements were
done for the two other plunger gates and the complete set of
$\alpha_\mathrm{PG}/\alpha_\mathrm{BG}$ is summarized in
Table\,\ref{leverarms}. These lever arms are consistent with the geometry of the sample (Fig.\,\ref{fig1}).

Fig.\,\ref{fig2}(e) shows a high resolution measurement
corresponding to a zoom of Fig.\,\ref{fig2}(d) (see the black square in Fig\,\ref{fig2}(d)). The narrow diagonal lines correspond to single Coulomb resonances in the quantum dot
and from the slope we determine $\alpha_\mathrm{PG}/\alpha_\mathrm{BG}$ for the dot. From
corresponding measurements varying the other two in-plane gates the
relative dot lever arms of all three plunger gates are extracted. These lever arms can also be found in Table\,\ref{leverarms}.

From Table\,\ref{leverarms} it can be seen that the relative lever arms with respect to the dot $\alpha_\mathrm{PG}^\mathrm{QD}/\alpha_\mathrm{BG}^\mathrm{QD}$ are very similar for all three plunger gates. However, the lever arms with respect to the different constrictions vary significantly. In particular the lever arm of each plunger gate with respect to the constriction on the opposite side of the dot is much weaker than all other lever arms. Hence, the plunger gate dependence of transport through the dot can be used to identify if changes are due to alterations of the dot wave function or the constriction resonances.


\subsection{Determination of individual conductances from the conductance matrix}

\begin{figure}
  \begin{center}
    \includegraphics[width=1\textwidth]{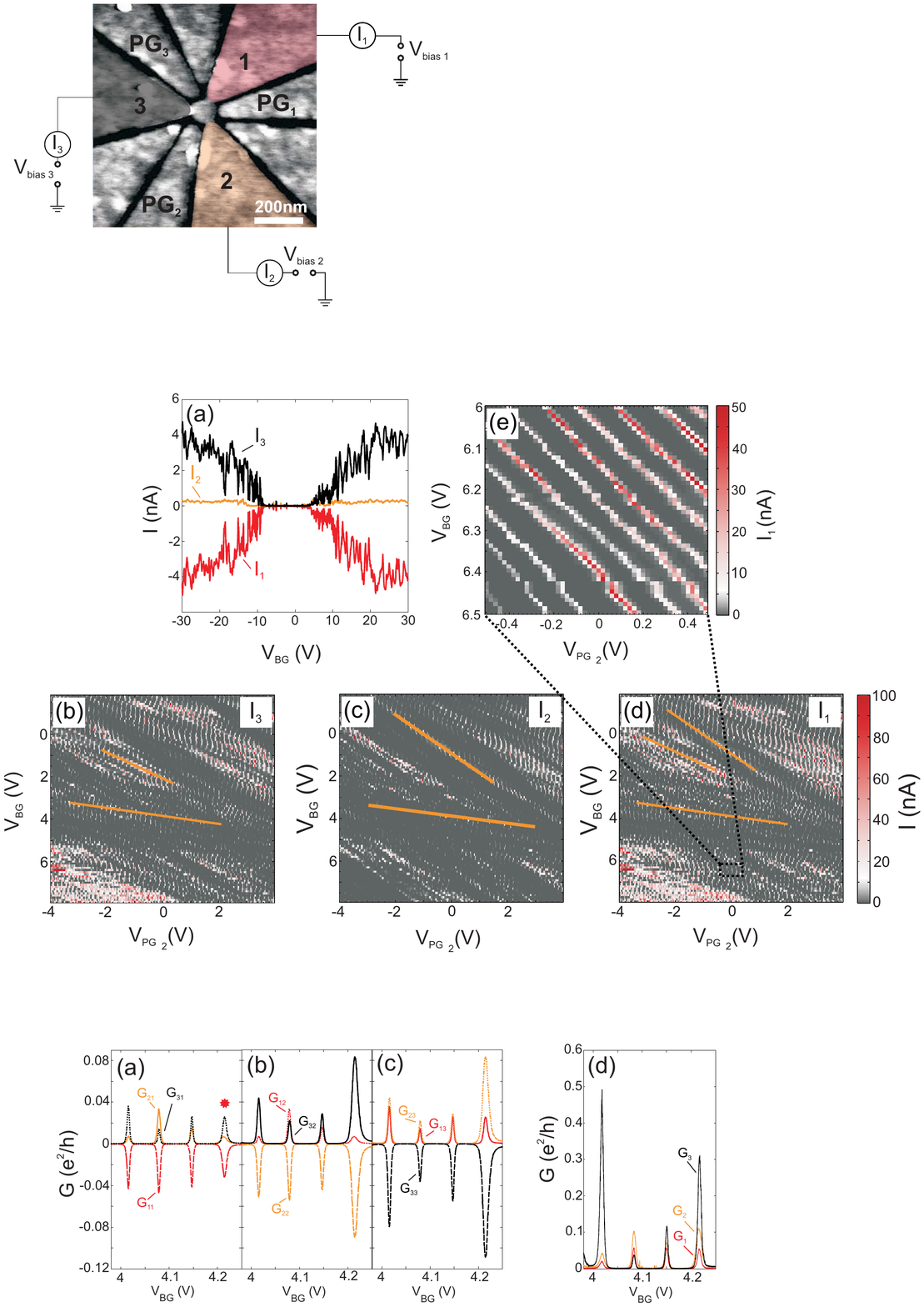}
    \caption{(a)-(c) Measurement of the complete conductance matrix for the system. The conductance in lead 1, 2 and 3 is plotted in red, orange and black respectively. The conductance in the biased lead is always plotted negative. (d) Corresponding individual conductances.}
    \label{fig3}
  \end{center}
\end{figure}

The conductance matrix $\mathbf{G}$ of a three-terminal system is
given by

\begin{equation}
    \left(
        \begin{array}{c}
            I_1\\
            I_2\\
            I_3
        \end{array}
    \right)
    =
    \left(
        \begin{array}{ccc}
            G_{11} & G_{12} & G_{13}\\
            G_{21} & G_{22} & G_{23}\\
            G_{31} & G_{32} & G_{33}
        \end{array}
    \right)
    \left(
        \begin{array}{c}
            V_1\\
            V_2\\
            V_3
        \end{array}
    \right)
    =
    \mathbf{G}
    \left(
        \begin{array}{c}
            V_1\\
            V_2\\
            V_3
        \end{array}
    \right).
\end{equation}
In Fig.\,\ref{fig3}(a)-(c) the nine elements of the conductance
matrix are shown, measured by applying a 100\,uV bias to lead 1, 2, and 3
respectively. There are two sum-rules that should be obeyed by the
conductance matrix. First, due to current conservation
$\Sigma_{i=1}^3 G_{ij}=0$ for all $j$. Second, if the same voltage
is applied to all leads no current should flow, $\Sigma_{j=1}^3
G_{ij}=0$ for all $i$. In addition, at zero magnetic field $\mathbf{G}$ should be symmetric, $G_{ij}=G_{ji}$. As a result there are only three independent conductance matrix elements, from which the complete matrix can be deduced. For the measurement shown in
Fig.\,\ref{fig3}(a)-(c) the first sum rule is obeyed with a
relative error less than 1\,\% of the highest current level, while
the second sum rule is obeyed with a relative error less than 10\,\%
of the highest current level. In order to obtain such a small error
the measurements are done very carefully. To
minimize the influence of voltage offsets in the measurement setup, measurements for positive and negative bias were averaged. In addition, to avoid errors due to instabilities of the sample all nine elements of
the conductance matrix for both positive and negative bias are
measured before each BG step. In other words, all conductance matrix elements are
measured within a single back gate sweep.

For single-level transport in the weak coupling regime the
individual tunnel couplings, $\Gamma$, between the dot and each lead
can be determined from the conductance matrix. \cite{beenakker1991} In this transport regime the width of the Coulomb peaks is peak-independent for a given temperature.
Looking at the Coulomb peaks in Fig.\,\ref{fig3}(a)-(c) it can be
seen that in our measurements the width of the peaks varies (especially pronounced for the forth peak). This is a sign of multi-level transport.



In the multilevel regime the
individual tunnel coupling strengths from the dot to each lead cannot be extracted directly from the
conductance matrix. However, if we consider the quantum dot as a
classical star-shaped conductance network, we can extract the three
individual conductances $G_k$ connecting lead $k$ to the dot from the
relation

\begin{equation}
    \mathbf{G}
    =
    \frac{1}{G_1+G_2+G_3}
    \left(
        \begin{array}{ccc}
            G_1(G_2+G_3) &      -G_1G_2 & -G_1G_3      \\
            -G_2G_1      & G_2(G_1+G_3) & -G_2G_3      \\
            -G_3G_1      &      -G_3G_2 & G_3(G_1+G_2) \\
        \end{array}
    \right).
\end{equation}

In Fig.\,\ref{fig3}(d), $G_1$, $G_2$ and $G_3$ obtained from the nine conductance matrix elements shown in Fig.\,\ref{fig3}(a)-(c), can be seen. The
individual conductances fluctuate largely from peak to peak demonstrating that the coupling strengths between the leads and the dot vary significantly from peak to peak.


\subsection{Temperature dependence}

\begin{figure}
   \begin{center}
    \includegraphics[width=1\textwidth]{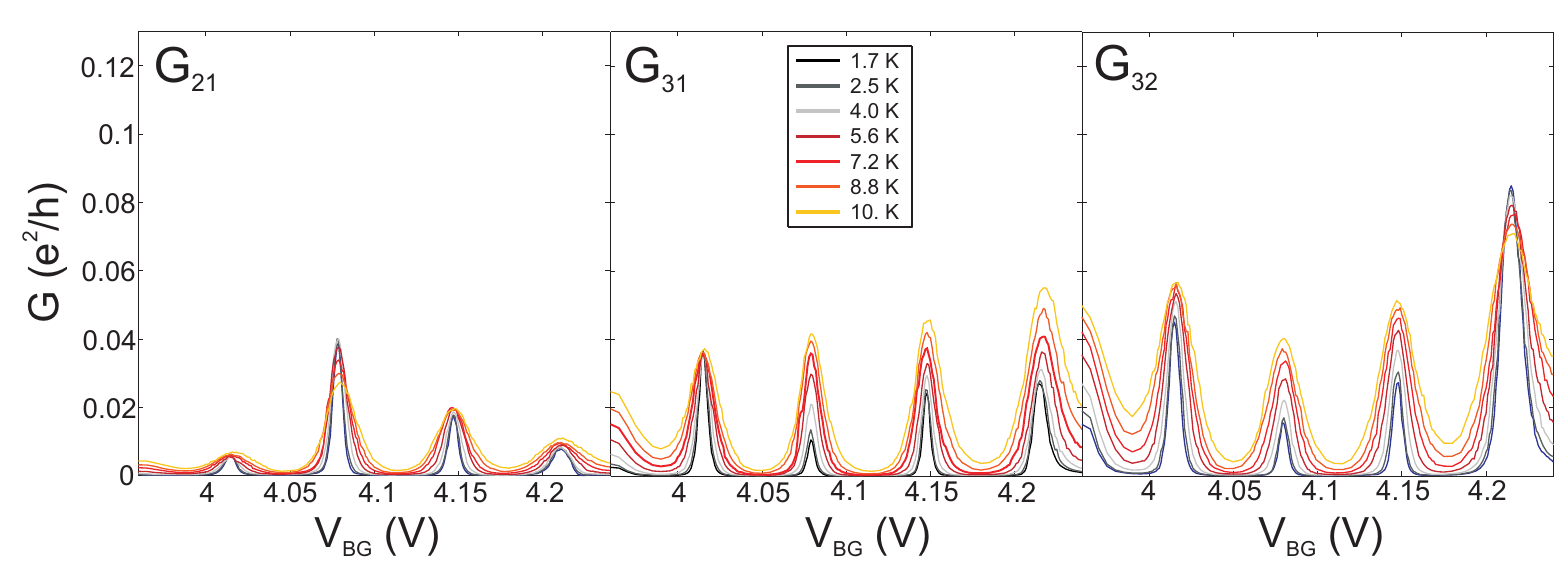}
    \caption{Temperature dependence of the four studied Coulomb peaks. The three independent matrix elements $G_{21}$, $G_{31}$ and $G_{32}$ are plotted. For the first and the third peak (from the left) the peak maxima increase with increasing temperature for all three conductance matrix elements. However, for the second peak the peak maxima increase with increasing temperature for $G_{31}$ and $G_{32}$ while it decreases for $G_{21}$. A similar behaviour with different temperature dependence for the different conductance matrix elements are seen for the fourth peak.}
    \label{fig4}
  \end{center}
\end{figure}

In the previous section signs of multi-level transport were seen. To further support this, we present the
temperature dependence of the Coulomb peaks shown in
Fig.\,\ref{fig3}. This is depicted in Fig.\,\ref{fig4} where the
three independent conductance matrix elements $G_{21}$, $G_{31}$ and $G_{32}$ are plotted as a function
of $V_\mathrm{BG}$ for seven different temperatures between 1.7\,K and 10\,K. In general it can be seen that all Coulomb peaks
broaden with increasing temperature. For the first peak (from the left) and
the third peak, the peak maxima increase for increasing temperature, which is a
signature of multilevel transport \cite{meir1991}. However, for the second peak the
peak maximum increases with increasing temperature for $G_{31}$ and
$G_{32}$ while it decreases for $G_{21}$. A similar behaviour is
seen for the fourth peak where the peak maximum increases with increasing
temperature for $G_{21}$ and $G_{31}$ while it decreases for
$G_{32}$. It should be noted that even though two peaks
are seen to decrease in height, they do not show the $1/T$-dependence as expected for true single-level transport.

It is known that in the multi-level regime the temperature dependence of Coulomb peaks can vary from peak to peak due to variations in the couplings between the leads and the different dot states. \cite{meir1991} However, the measurement of different temperature dependences of conductances measured in different leads for the same Coulomb peaks is unique to a three or more terminal system and has to our knowledge not been measured before. Following the arguments of Ref.\,\cite{meir1991}, our results suggest that the different leads couple with different strengths to the different dot states involved in transport. This is further supported by detailed measurements of single Coulomb peaks as discussed below.


\subsection{Shift between Coulomb resonance positions due to coupling of
different leads to different dot states}
\label{shift}

\begin{figure}
  \begin{center}
    \includegraphics[width=0.6\textwidth]{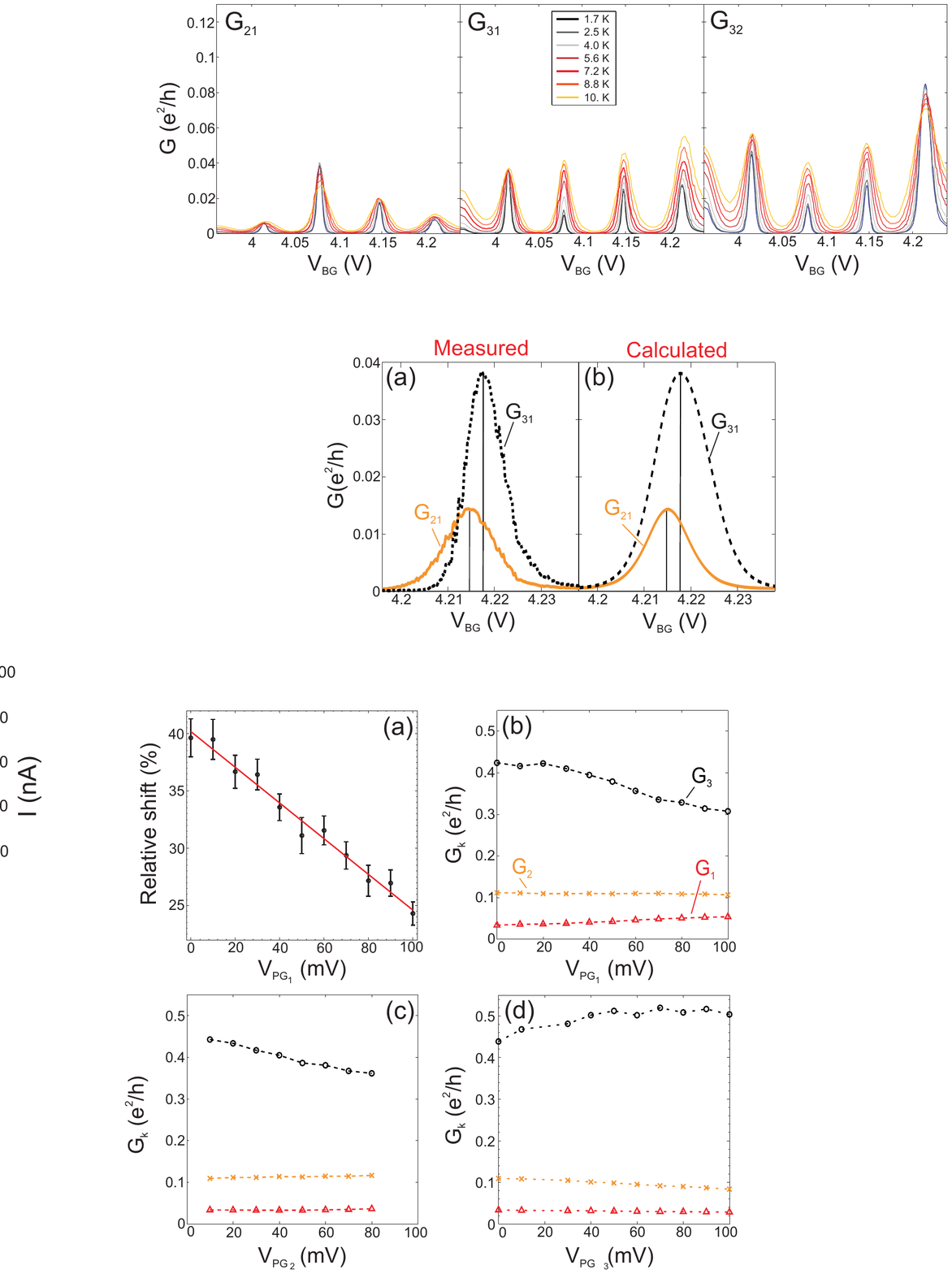}
    \caption{(a) Measurement of $G_{21}$ and $G_{31}$ as a function of
    V$_\mathrm{BG}$ for a single Coulomb peak. The maxima of the two
    peaks are shifted by 2\,mV in back gate voltage corresponding to 0.9\,$k_\mathrm{B}T$ when assuming that the FWHM of the peaks are $\approx$ 4.4\,$k_\mathrm{B}T$.
    (b) Corresponding calculation with a two-level model where the three leads
    couple differently to the two dot levels.}
    \label{fig5}
  \end{center}
\end{figure}

When studying single Coulomb peaks in detail we frequently observe that
peaks corresponding to conductances in different leads have their maxima
at slightly different positions in gate-voltage. An example of this is shown
in Fig.\,\ref{fig5}(a) where $G_{31}$ and $G_{21}$ are plotted for the fourth peak in Fig.\,\ref{fig3}(a) (see red star). A 100\,$\mu$V bias voltage is applied to lead 1, a voltage of 100\,mV is applied to gate 1 while all other gates are grounded. The conductance $G_{31}$ (black dotted curve) has its
maximum at 4.217\,V while $G_{21}$ (orange solid curve) has its
maximum at 4.215\,V. For multi-level transport the full width at half maximum (FWHM) of a Coulomb peak is $\approx 4.4\,k_\mathrm{B}T$, where $k_\mathrm{B}$ is the Boltzmann constant and $T$ is the temperature \cite{beenakker1991}. From the FWHM of the measured peaks we then estimate the shift between the maxima of $G_{31}$ and $G_{21}$ to be 0.9\,$k_\mathrm{B}T$.

It should be noted that the measurements shown in Fig.\,\ref{fig5}(a) are not averaged between conductances measured for positive and negative bias (unlike in Fig.\ref{fig3}). The gating effect of the lead where the bias is applied shifts the Coulomb resonances in energy at finite bias voltages. For positive bias voltages the resonances are shifted to more positive back gate voltages, while for negative bias voltages they are shifted to more negative back gate voltages. Averaging would therefore result in a broad resonance with a maximum positioned between the maxima of the original resonances. By only considering the conductances measured with the bias voltage applied to the same lead (for positive or negative bias voltage), the gating effect of the source causes the same shift for all resonances. Hence, the shift of 2\,mV between the maxima of $G_{21}$ and $G_{31}$ found above is not due to the gating effect of the source.

The shift between the maxima of the conductances of $G_{21}$ and $G_{31}$ can be understood in terms of
multi-level transport where the leads couple with different strengths to the different dot states.  In order to illustrate this effect qualitatively, we calculate $G_{31}$ and $G_{21}$ for the simplest possible multi-level system, a two level system. We use the rate equation approach introduced by Beenakker in Ref.\,\cite{beenakker1991} extended to a three-terminal dot with two levels contributing to the current.

In this model it can be shown that the current in each lead is the sum of two contributions, one current via the first level and one current via the second level.
The shift between currents in two different leads is determined by three parameters, the single particle level spacing $\Delta$ and two parameters determining how the current in each lead is distributed between the two dot levels. The measured shift can be qualitatively reproduced by the model for a large range of parameters. An example of a calculation showing good agreement with the experiment in Fig.\,\ref{fig5}(a) is depicted in Fig.\,\ref{fig5}(b). In order to put more constraints on the values of the parameters we also tried to reproduce the temperature dependence of the peaks. Unfortunately, with a two level model it is not possible to quantitatively reproduce the observed shift and the observed temperature dependence at the same time. Thus, we here most likely have more than two levels involved in transport. Still, we would like to emphasize that the simple two-level model does qualitatively reproduce the shift, supporting that it is indeed due to the different coupling of different leads to different dot states. This also agrees with the interpretation of the temperature dependence of the Coulomb peaks as discussed above.


\subsection{Evolution of Coulomb resonance shift with in-plane gate voltages}

\begin{figure}
  \begin{center}
    \includegraphics[width=0.8\textwidth]{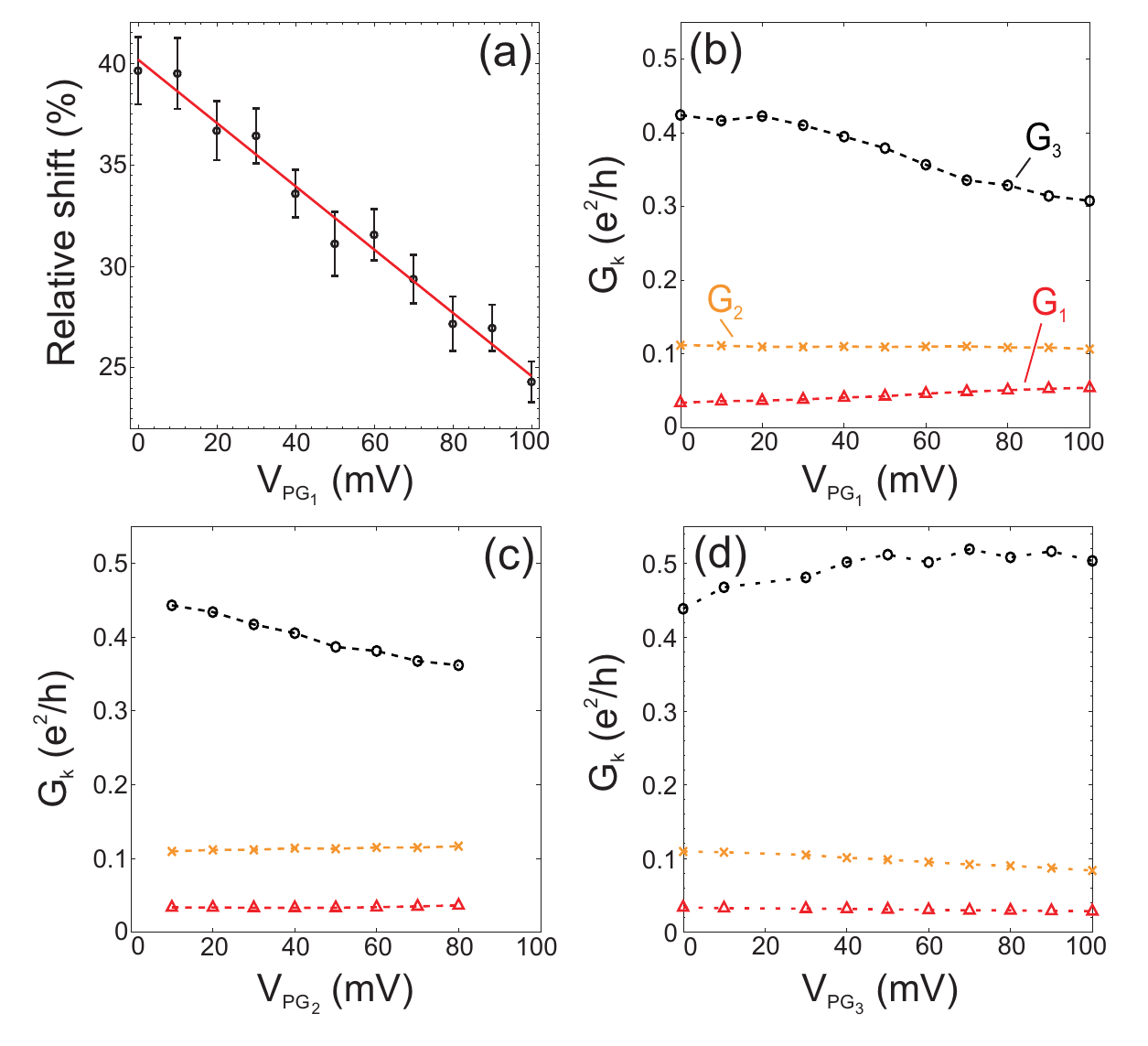}
    \caption{(a) Shift relative to the average FWHM of the peaks as a function of $V_{\mathrm{PG}_1}$. The red line is a linear fit to the data points. (b) Corresponding evolution of the individual conductances as a function of $V_{\mathrm{PG_1}}$. $G_2$ hardly changes, $G_1$ changes only slightly, while $G_3$ changes significantly. For the evolution of the individual conductances as a function of $V_{\mathrm{PG}_2}$ (c) and $V_{\mathrm{PG}_3}$ (d) $G_1$ and $G_2$ again change only slightly while $G_3$ changes significantly. }
    \label{fig6}
  \end{center}
\end{figure}

From the above discussion we argue that the observed shift is due to different leads coupling with different strengths to different quantum dot states. However, it is still an open question if small changes in the coupling strengths between the leads and the dot are dominated by changes of the localized states in the constrictions or changes in the dot wave function.

Fig.\,\ref{fig6}(a) shows the shift from Fig.\,\ref{fig5}(a) relative to the average FWHM of the peaks as a function of $V_{\mathrm{PG}_1}$. It can be seen that already for small plunger gate voltages the shift can be tuned significantly and systematically by an in-plane gate. Fig.\,\ref{fig6}(b) shows the evolution of the individual conductances $G_k$ of the three leads with $V_{\mathrm{PG}_1}$. Fig.\,\ref{fig6}(c) and (d) show the corresponding evolution of $G_1$, $G_2$ and $G_3$ for $V_{\mathrm{PG}_2}$ and $V_{\mathrm{PG}_3}$ respectively. If the changes in the coupling between the dot states and the lead states would be dominated by changes in the localizations in the constrictions we would expect a correlation between the evolution of the individual conductances and the relative lever arms of the plunger gates with respect to the constrictions (see Table\,\ref{leverarms}). However, no correlation is found. $G_3$ is influenced the most by all three plunger gates. $G_2$ and $G_1$ are only changed slightly. We therefore conclude that the Coulomb blockade resonances and in particular the amplitudes of the current maxima investigated here are mostly governed by the wave function in the dot and to a lesser extent by localization sites in the leads.

Furthermore, it can be seen that in general the effect of PG$_3$ is opposite to the effect of PG$_2$ and PG$_1$. This might indicate that it is indeed not random how the $G$'s are changing and that such measurements could be used to obtain further qualitative understanding on how the dot wave function is distributed in the dot. The lack of any geometric correlation with the evolution of the $G$'s also suggests that the plunger gates tune the dot wave function as a whole, rather than several independent puddles.


\section{Conclusion}

We have investigated a three-terminal graphene quantum dot in the multi-level Coulomb blockade regime. The dot was thoroughly characterized by both gate-gate sweeps where all relative lever arms could be extracted, and temperature dependent measurements. When investigating single Coulomb peaks in more detail a shift in peak maxima between conductances measured in the different leads were observed. This result can be qualitatively reproduced by a rate equation model where different leads couple differently strong to different dot states. The shift can be tuned by the plunger gates and by investigating the corresponding evolution of the individual conductances we find no correlation between this evolution and the relative lever arm determined. We therefore conclude that on small energy scales the changes in coupling are due to changes in the dot wave function, which is rather a single wave function extended over the dot than several localized states. This is an important insight in view of the potential to use graphene quantum dots for spin qubits.


\section*{Acknowledgements}

Financial support from the Swiss Science Foundation is gratefully acknowledged.



\end{document}